\documentclass[a4paper, 11pt]{article}

\usepackage{graphics}
\usepackage{epstopdf}
\usepackage{float}
\usepackage{paralist}

\usepackage[latin1]{inputenc}
\usepackage{lscape,url}
\usepackage{graphicx}
\usepackage{amsmath}
\usepackage{amssymb}
\usepackage{amsfonts}
\usepackage{amssymb}
\usepackage{anysize}
\usepackage{color}

\usepackage{fancyhdr}
\usepackage{latexsym}
\usepackage{array}
\usepackage{multirow}
\usepackage{textcomp}

\begin{document}


\begin{center} 
{\LARGE A nonlinear optimisation model for constructing minimal drawdown portfolios} \\
~\\
C.A. Valle$^1$ and J.E. Beasley$^2$ \\
~\\
$^1$Departamento de Ci\^encia da Computa\c c\~ao, \\ 
Universidade Federal de Minas Gerais, \\ Belo Horizonte, 
MG 31270-010, Brazil \\
arbex@dcc.ufmg.br
\\
$^2$Mathematical Sciences,
Brunel University,
Uxbridge UB8 3PH, UK \\
 john.beasley@brunel.ac.uk \\
~\\
August 2019 \\

\end{center}

\begin{abstract}
In this paper we consider the problem of minimising drawdown in a portfolio of financial assets. Here drawdown represents the relative opportunity cost of the single best missed trading opportunity over a specified time period. We formulate the problem  (minimising average drawdown, maximum drawdown, or a weighted combination of the two) as a nonlinear program and show how it can be partially linearised  by replacing one of the nonlinear constraints by equivalent linear constraints. 

Computational results are presented (generated using the nonlinear solver SCIP) for three test instances drawn from the EURO~STOXX~50, the FTSE~100 and the S\&P~500 with daily price data over the period 2010-2016. 
We present results for long-only drawdown portfolios as well as results for portfolios with both long and short positions.
These indicate that (on average) our minimal drawdown portfolios dominate the market indices in terms of return, Sharpe ratio,     maximum drawdown and average drawdown
 over the (approximately 1800 trading day) out-of-sample period.

\end{abstract}

\textbf{Keywords:} Index out-performance; Nonlinear optimisation; 
 Portfolio construction; Portfolio drawdown; Portfolio optimisation

\section{Introduction}

Given a portfolio of financial assets then, as time passes, the price of each asset changes and so by extension the  value of the portfolio changes. Portfolio drawdown is a measure of current portfolio value when compared to the maximum value achieved by the same portfolio of assets in the recent past. It gives insight into how much the portfolio has fallen in value by comparing its value now with the best (maximum) value it had in the recent past.

In this paper we adopt an optimisation approach to the problem of deciding a portfolio that minimises portfolio drawdown. The structure of this paper is as follows. In 
Section~\ref{sec:lit} we give an example to illustrate the concept of drawdown and
then review the relevant literature relating to deciding portfolios that minimise drawdown. We also discuss the context of our work, Operational Research applied to a financial portfolio optimisation problem. 
In 
Section~\ref{sec:form} we give a nonlinear formulation of the problem of 
deciding a portfolio that minimises drawdown. Our formulation incorporates cash inflow/outflow, and can be used either to create an initial portfolio from cash or to rebalance an existing portfolio. Transaction costs associated with buying or selling an asset are included. 
We indicate how we can partially  linearise our formulation.
We also discuss some computational issues associated with our nonlinear formulation and present the amendments necessary to deal with shorting. 
In Section~\ref{sec:results} we give computational results for constructing minimum drawdown portfolios for three different problem instances 
derived from universes defined by major equity markets, involving up to 500 assets. 
We present results for long-only drawdown portfolios as well as results for portfolios with both long and short positions. 
Finally in 
Section~\ref{sec:con} we present our conclusions.

\section{Drawdown example and literature review}
\label{sec:lit}

In this section we first give an example to illustrate the concept of drawdown. We then present our literature review relating to deciding portfolios that minimise drawdown. We also discuss the context of our work to show that it follows the general pattern seen in the literature for Operational Research applied to financial portfolio optimisation problems.

\subsection{Drawdown illustration}

\begin{figure}[!htb]
\begin{center}
\caption{Drawdown illustrated}\label{fig1}
\includegraphics[width=0.9\textwidth]{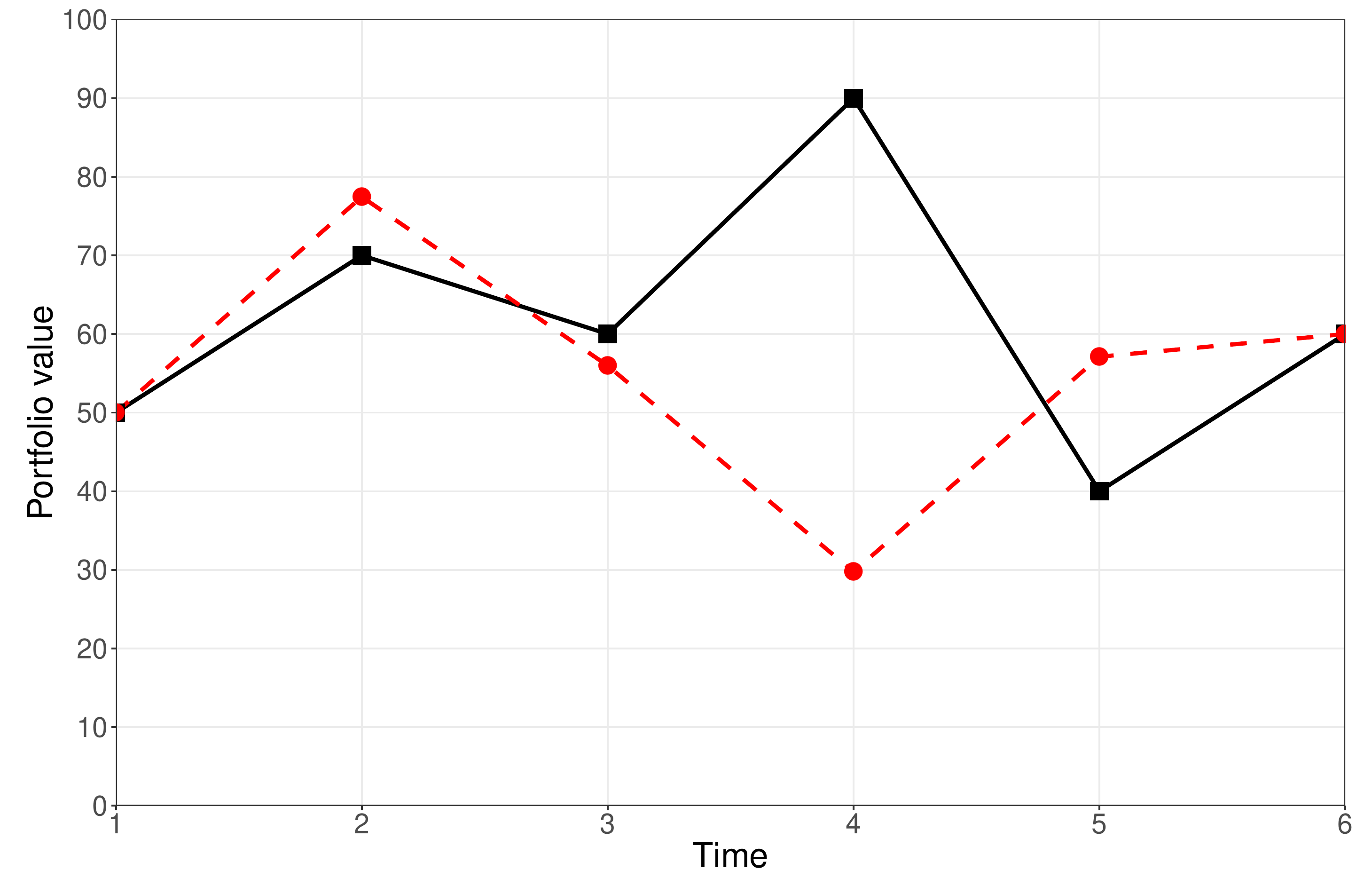}
\end{center}
\end{figure}

To illustrate drawdown consider the solid line in 
 Figure~\ref{fig1} where we show the value of a portfolio over time (starting from a  value of 50 at time one). At time 6 the portfolio has value 60 and we can see that the maximum value of the portfolio was 90 (at time 4). At time 6 therefore the drawdown associated with this portfolio can be defined as $100(90-60)/90 = 33.33\%$, i.e.~as the reduction in portfolio value, from the best (maximum) portfolio value seen over the time period considered, expressed as a percentage of maximum portfolio value. 

In a similar manner drawdown at time 5 is $100(90-40)/90 = 55.56\%$, at time 4 is zero, at time 3 is $100(70-60)/70 = 14.29\%$, at times 2 and 1 is zero. The maximum drawdown over the entire period is that associated with time 5, so 55.56\% and the mean drawdown over the entire period is 17.20\%.

Note here that a portfolio which always increases in value (over the time horizon considered) would have a drawdown of zero for each and every time period. 

Table~\ref{table1} shows (to two decimal places) the value over time of the solid portfolio in Figure~\ref{fig1}, the single period (continuous, logarithmic)  return and the drawdown.

\begin{table}[!htb]
{\scriptsize
\caption{Portfolio drawdown examples}\label{table1}
\begin{tabular}{cccccccc}
\hline
	&	\multicolumn{3}{c}{Solid portfolio}			& &	\multicolumn{3}{c}{Dotted portfolio} \\  
	\cline{2-4} \cline{6-8}
Time	&	Value	&	Return (\%)	&	Drawdown (\%)	& &	Value	&	Return (\%)	&	Drawdown (\%)	\\   
 \hline

1	 &	50	 &		 &	0	 &	 &	50	 &		 &	0	 \\ 	   
2	 &	70	 &	33.65	 &	0	 &	 &	77.45	 &	43.76	 &	0	 \\ 	   
3	 &	60	 &	-15.42	 &	14.29	 &	 &	55.97	 &	
-32.48	 &	27.73 \\ 	   
4	 &	90	 &	40.55	 &	0	 &	 &	29.76	 &	-63.17	 &	61.58	 \\ 	   
5	 &	40	 &	-81.09	 &	55.56	 &	 &	57.11	 &	65.18	 &	26.26	 \\ 	   
6	 &	60	 &	40.55	 &	33.33	 &	 &	60	 &	4.94	 &	22.53 \\ 	   
\hline 	   
Mean	 &		 &	3.65	 &	17.20 &	 &		 &	3.65	 &	23.02	 \\ 	   
Standard deviation	 &		 &	52.84	 &		 &	 &		 &	52.84	 &		 \\ 	 

	\hline	 
											 
\end{tabular}
}
\end{table}

The value of drawdown at any particular time $t$ is (in relative
percentage terms) the opportunity cost of  the single best missed trading opportunity associated with selling the portfolio at some point before $t$ and then repurchasing it at $t$. It represents the (percentage) value forgone by not having sold the portfolio at its previous maximum value point 
(the single best  point at which to sell the portfolio)
and then repurchasing it at time $t$.

So in the example considered above we start at time 1 with a portfolio of assets. At time 6 had we sold the portfolio at time 4 (when its value was at a maximum) and then repurchased it at time 6 we would have banked 90 in cash from the sale of the portfolio
at time 4, and would need to spend 60 at time 6 to repurchase the portfolio. So at time 6 we would be in the same situation as we started at time 1, holding the portfolio of assets, but now also   with 30 in cash. Since we did not avail ourselves of this trading opportunity our percentage missed profit is $100(30)/90 = 33.33\%$, i.e.~the drawdown value at time 6.
As is usual in the literature the drawdown value given here does not take into consideration 
any transaction costs that may have been incurred 
  if the missed trading opportunities had been actioned.

As this missed profit example indicates measuring drawdown is of interest to funds that essentially hold portfolios for short-term profit. In practical financial portfolio management drawdown is of especial interest to hedge funds and commodity futures 
funds (also known  as
Commodity Trading Advisors, CTA's)~\cite{goldberg16,lhabitant09}. In the USA, for example, the Commodity Futures Trading Commission mandates that drawdown must be disclosed.
Many hedge funds have a relatively short-term time horizon and seek to make profit in the short-term and so as an internal performance measure would calculate drawdown (irrespective as to whether it needs to be publicly disclosed to investors or not).

For a fund such as a hedge fund the value of drawdown may factor into the decision by individual investors as to whether (or not) to withdraw their investment from the fund when they have the opportunity to do so.  
Investors in hedge funds are often looking for short-term profit and so have a natural expectation that such funds would use the expertise that they claim to possess to take advantage of trading opportunities. 
As discussed above drawdown is a measure that provides a numeric value for a missed trading opportunity and hence drawdown gives investors some insight into how their investments have been managed. Clearly for a hedge fund their income relies on their ability to both attract and retain investment from their investors.

Readers familiar with financial portfolio management will be aware that often portfolios are examined/chosen in terms of their mean return and variance in return. The classic mean-variance portfolio optimisation approach due to Markowitz~\cite{markowitz52} would be the archetypal example of this. 
Burghardt and Walls~\cite{burghardt11} make the point that two portfolios with identical mean returns and identical variances (equivalently identical standard deviations) in returns can have very different drawdown values. 

As an illustration of this the dotted line shown in Figure~\ref{fig1} shows  another portfolio that also has value 50 at time 1 and value 60 at time 6, but a different behaviour over time. 
 It can be clearly seen from Table~\ref{table1} that the two portfolios shown in Figure~\ref{fig1} have identical mean returns and standard deviations in returns, but very different drawdown values. 

Drawdown is defined based on the \textbf{\emph{sequence}} of values achieved,  in other words it is a \textbf{\emph{path-dependent}} measure. This contrasts with path/sequence
 independent summary statistics such as mean and variance/standard deviation
calculated from an entire set of values. For this reason any permutation of the set of returns associated with either of the two portfolios shown in Table~\ref{table1} will result in different portfolios with different drawdown values (but all such portfolios will have value 50 at time 1, value 60 at time 6, the same set of returns, the same mean return and the same standard deviation in return).

\subsection{Literature review}

To structure our literature review we first discuss papers that consider drawdown within a discrete time setting. In such papers financial asset values are assumed to be known, e.g.~from historic data, at discrete points in time. This is a standard setting for portfolio optimisation such as adopted in  mean-variance optimisation~\cite{markowitz52}.  

We then discuss papers that use a continuous time setting. Such papers typically start from the premise that the process underlying asset price dynamics is some form of Brownian motion (e.g.~geometric Brownian motion with drift) and  involve stochastic differential equations. 

We exclude from explicit consideration here a number of papers which, whilst using drawdown as a statistic with which to evaluate portfolio performance, have as their primary focus other matters.

\subsubsection{Discrete time}

Alexander and Baptista~\cite{alexander06} introduced a drawdown constraint into the standard mean-variance portfolio approach due to  Markowitz~\cite{markowitz52}. Their approach was a scenario based one in which asset returns in each of a number of different scenarios are known. They defined drawdown as the worst (minimum) portfolio return seen over all the scenarios considered. 
With drawdown defined in this way limiting drawdown merely involves adding linear constraints (one for each scenario considered) to the standard quadratic program associated with mean-variance portfolio optimisation.  
Using one data set containing ten assets they illustrated how the standard unconstrained mean-variance efficient frontier compared with the drawdown constrained efficient frontier.

Yao et al~\cite{yao13} extended the work of 
Alexander and Baptista~\cite{alexander06} from a  theoretical standpoint. They investigated the composition and geometric features of the mean-variance efficient frontier with a drawdown constraint. 
One numeric example involving eight assets was presented.

Baghdadabad et al~\cite{baghdadabad13} linked drawdown to the 
Capital Asset Pricing Model 
(CAPM)~\cite{lintner1965, sharpe1964, treynor1961}. 
In the CAPM the setting is that different assets achieve different returns, against a common background of a risk-free rate and a market return. The hypothesis is that the  return on any asset in excess of the risk-free rate is a linear function, namely 
a constant (alpha) plus
a constant (beta) multiplied by the difference between the return on the market and the risk-free rate. Values for alpha and beta, which are typically different for different assets, can be obtained from data using standard linear regression.
Drawdown is defined in~\cite{baghdadabad13} as the loss in portfolio value when compared with a previous maximum value. 
They presented two beta values based upon maximum drawdown and average drawdown. These were computed using the covariance of a function that involves the sum of asset drawdown and asset return above/below its mean. Extensive results were presented for their beta values when computed for 11737 mutual funds over the  period 2000-2011.

Zabarankin et al~\cite{zabarankin14} also linked drawdown to the 
CAPM. Their approach is fundamentally different from that of
Baghdadabad et al~\cite{baghdadabad13}.
They considered
Conditional Drawdown-at-Risk (CDaR),
which is defined as the average of a specified percentage of the largest drawdowns over an investment
horizon.  Their approach was a scenario (sample-path) based approach in which asset returns in each of a number of different scenarios were known (as were the probabilities of each scenario). 
Drawdown was defined using cumulative portfolio return since the initial investment and was equal to the difference between the maximum cumulative portfolio return over a specified number of preceding periods and the current portfolio cumulative return. 
They proposed either minimising portfolio CDaR subject to a constraint on portfolio expected return at the end of the investment horizon; or  maximising portfolio expected return at the end of the investment horizon subject to a constraint on portfolio CDaR. 
Underlying their work is the earlier work presented in the literature~\cite{chekhlov03,chekhlov05}
 relating to the definition of conditional drawdown.
The main focus of their paper was on the CAPM, with CDaR alpha and CDaR beta being defined, these being analogous to alpha and beta in the CAPM. Values for CDaR alpha and CDaR beta were computed and presented for 80 hedge fund indices.

Mohr and Dochow~\cite{mohr17} considered how expert judgment can be incorporated into portfolio selection so as to minimise maximum drawdown. In their work drawdown was defined as the difference between the maximum portfolio value to date and current portfolio value.
Computational results were presented for portfolios containing just two assets formed
from a NYSE data set containing daily price data for 36  assets over 
5651 days until the end of 1985.

\subsubsection{Continuous time}

Grossman and Zhou~\cite{grossman93} considered the case 
of an investor who wants 
to lose no more than a fixed percentage of the maximum value their portfolio has achieved to date. This constraint is equivalent to saying that drawdown (defined in percentage terms, as in the example considered in Table~\ref{table1} above) 
in any period is limited. For a portfolio comprising just one risky asset and a risk-free asset they derive an optimal investment policy.

Magdon-Ismail and Atiya~\cite{ismail2004} presented results relating maximum drawdown to the mean return and 
Sharpe ratio~\cite{sharpe66,sharpe75,sharpe94}. Related work is presented in
Magdon-Ismail et al~\cite{ismail2004a} who considered the distribution of drawdown and its expected value. 

Pospisil and  Vecer~\cite{pospisil10} considered a financial instrument whose value depends not only on the price of an underlying asset, but also on other factors such as running maximum drawdown. They examined the  
sensitivity of the value of this instrument with respect to drawdown.
Their work draws on earlier work presented by the same authors 
in~\cite{pospisil08}. In their work drawdown was defined as the difference between the maximum portfolio value to date and current portfolio value. 

Yu et al~\cite{yu14} build on the work of 
Grossman and Zhou~\cite{grossman93} and 
considered the case of deriving the optimal investment policy for a portfolio comprising of a number of risky assets and a risk-free asset. Numeric results  for a single risky asset and for two risky assets were presented.

Chen et al~\cite{chen15} considered two cases: a portfolio with two risky assets and a portfolio with one risky asset and a risk-free asset. They focused on minimising the probability that a significant drawdown occurs over portfolio lifetime. In their work drawdown is defined in percentage terms, as in the example considered in 
Table~\ref{table1} above, and a significant drawdown is one that exceeds a specified percentage. 
Related work can  be seen in Angoshtari et al~\cite{angoshtari16}
for a portfolio with a payout rate, one risky asset and a risk-free asset and in Angoshtari et al~\cite{angoshtari16a} for a portfolio with a constant payout rate, one risky asset and a risk-free asset.

Goldberg and Mahmoud~\cite{goldberg16} considered Conditional Expected Drawdown (CED), which is the expected value of maximum drawdown given that a specified maximum drawdown threshold has been exceeded. CED is the tail mean of the distribution of maximum drawdowns. Drawdown was defined as the difference between the maximum portfolio value to date and current portfolio value. 
Although set within a continuous time framework their work does not depend on restrictive assumptions as to the underlying stochastic process. They presented CED values for a US equity index, a US bond index and three fixed-mix portfolios using daily data over the period 1982-2013.

Mahmoud~\cite{mahmoud17} considered drawdown as a temporal 
path-dependent risk measure within a stochastic continuous time framework. 
 In their work drawdown was defined as the difference between the maximum portfolio value to date and current portfolio value. 
Although this paper is primarily theoretical in nature drawdown associated with two US equity and bond indices over the period 1978-2013 is presented for illustration.

\subsection{Context}
The basic context of our work is financial portfolio optimisation from an Operational Research (OR) perspective. Here the financial problem is to decide a portfolio of financial assets (i.e.~decide how much to invest in each asset) so as to optimise some objective subject to constraints upon the investments made. The constraints applied might limit the investment made in any particular asset as well as specify the total investment that is to be made.

In dealing with financial portfolio optimisation from an OR perspective portfolio choice is underpinned by a formal explicit optimisation model, which is solved either optimally or heuristically (e.g.~using some metaheuristic). Typically a heuristic is adopted when the underlying optimisation model is (computationally) hard to solve, for example when it involves integer variables and/or nonlinearities.

For any particular financial portfolio optimisation problem the general pattern  is that it first appears in the  academic literature in journals related to the financial sphere (as would seem natural). A number of years after its first appearance in the financial literature  it  captures the attention of one or more OR workers and the problem then appears in journals more commonly associated with OR. This general pattern is illustrated and evidenced below with reference to three example financial portfolio optimisation problems.\\

\noindent \textbf{\emph{Markowitz mean-variance portfolio optimisation with cardinality constraints}}

Here the problem is to find a portfolio that best balances risk against return, where risk is measured using variance in portfolio return, but where in addition there are constraints on the number of assets that can be in the portfolio chosen. This is the classic mean-variance portfolio optimisation approach due to 
Markowitz~\cite{markowitz52}, but enhanced with cardinality constraints. 

One early work in the finance literature formalising the problem of restricting the number of assets in a mean-variance portfolio is that of Jacob~\cite{jacob74} which appeared in a finance journal in 1974. Early work in the literature from an OR perspective in OR journals  can be found in~\cite{ bienstock96, chang2000, maringer03}. That work dates from 1996 some 22 years after appearance  of the problem in the financial literature. 
Recent work in  OR journals relating to the problem 
includes~\cite{andriosopoulos19, cesarone13, gao13, guijarro18, hardoroudi17, lee19,
liagkouras18, liagkouras18a, liu16, xidonas18, 
zhou18}. \\

\noindent \textbf{\emph{Equity index tracking}}

Here the problem is to find a portfolio that best matches the return on a specified equity market index, but without holding all of the assets that are present in the index (so excluding full replication of the index).

Early work in the finance literature dealing with equity index tracking can be found in Rudd~\cite{rudd80}, which 
 appeared in a finance journal in 1980.
Early work in the literature from an OR perspective in OR journals  can be found in~\cite{adcock94, beasley03}. That work dates from 1994 some 14 years after appearance  of the problem in the financial literature.
Recent work in  OR journals relating to the problem includes~\cite{ andriosopoulos19, chen12,  giuzio16, mezali13, santa17, scozzari13, wu14}. \\

\noindent \textbf{\emph{Enhanced indexation (enhanced index tracking)}}

Here the problem is to find a portfolio that exceeds the return on a specified equity market index. 
 
Early published work in the finance field dating from
from the late 1990's dealing with enhanced indexation can be found in~\cite{ gold01, riepe98, scowcroft03}. Early work in the literature from an OR perspective in an OR journal  can be found in~\cite{canakgoz09}. That work dates from 2009 some 11 years after appearance  of the problem in the financial literature.
Recent work in  OR journals relating to the problem 
includes~\cite{ bruni17, filippi16, guastaroba16, mezali13, roman13, zhao19}. \\

\noindent Note here that for the three financial portfolio optimisation problems considered above it took over a decade in each case before the problem first captured the attention of OR workers. 

 The problem considered in this paper of finding a minimal drawdown portfolio, \textbf{\emph{follows exactly the same pattern as evidenced above}}. This problem, as seen in the detailed literature survey presented, has first appeared in the financial literature. The work presented in this paper is one of the first to consider the problem from an OR perspective. We believe that the financial portfolio optimisation problem relating to  minimal drawdown 
considered in this paper will  increasingly attract the attention of OR workers.

\section{Formulation}
\label{sec:form}

In this section we  formulate the problem of deciding a portfolio that minimises drawdown. 
Our formulation incorporates cash inflow/outflow, and can be used either to create an initial portfolio from cash or to rebalance an existing portfolio. Transaction costs associated with buying or selling an asset are included.

Our formulation involves a nonlinear definition of drawdown. 
We indicate how we can partially  linearise our formulation
 by replacing one of the nonlinear constraints by equivalent linear constraints.
We also discuss some computational issues associated with our nonlinear formulation.
We present the amendments to the formulation necessary to deal with shorting.

The approach adopted in formulating the problem below is a \textbf{\emph{backward-looking future-blind}} approach. Here we assume that we are at time $T$, with no knowledge of future asset prices, but with information as to historic asset prices over the 
time period $[1,T]$. This contrasts with a \textbf{\emph{forward-looking future-assumed-known}} simulation style approach of generating one or more sample paths (scenarios) for future asset prices and using that sample path information to decide a portfolio, e.g.~\cite{alexander06, yao13, zabarankin14}.

\subsection{Preliminaries}
Suppose we observe over time $ 1, 2, \ldots, T$ the value (price) of $N$ assets. Given this information the decision problem we face is how can we best invest at time $T$ in a portfolio which, had we held it over $[1,T]$, would best minimise an appropriate objective involving drawdown. To introduce our notation, let 
\begin{table}[H]
\begin{tabular}{p{1.3cm}p{13.8cm}}

$D$ & be the number of  time periods associated with calculating drawdown  \\
$P_t$ & be the value ($\geq 0$) of the portfolio at time $t$ \\
$d_t$ & be the portfolio drawdown value ($\geq 0$) at time $t$ \\
$M_t$  & be the maximum portfolio value ($\geq 0$) over the period $[ t, t-1, \ldots, max[1,t-D]]$ \\

\end{tabular}
\end{table}

Suppose the current time is $t$. Then the portfolio currently has value $P_t$. In the immediate $D$ time periods preceding time $t$ the portfolio had values $P_{\tau}, {\tau}=t-1, \ldots, max[1,t-D]$. At time $t$ therefore the drawdown associated with the portfolio  is given by:
\begin{equation}
M_t = max[~P_{\tau} ~|~{\tau}=t, t-1, \ldots, max[1,t-D]]  
\;\;\; t = 1, \ldots, T
\label{eq1b}
\end{equation}
\begin{equation}
d_t = 100(M_t - P_t) /M_t
\;\;\; t = 1, \ldots, T
\label{eq1}
\end{equation}

Equation~(\ref{eq1b}) defines the maximum portfolio value seen over the time period \\
 $[ t, t-1, \ldots, max[1,t-D]]$.
Equation~(\ref{eq1})  defines drawdown as the reduction in portfolio value  from the best portfolio value, expressed as a percentage of $M_t$.
Defining drawdown in relative percentage terms as in 
equation~(\ref{eq1}) is common in the literature, especially when reporting on the performance of a portfolio, e.g.~see~\cite{
allen16,
boigner15,
burghardt11, 
lhabitant09, 
madhogarhia15}.

The use of $\tau=t$ in the maximization term involving $P_{\tau}$ in 
equation~(\ref{eq1b})
 is deliberate and ensures that if the current value $P_t$ is superior to the values achieved in the preceding $D$  time periods then drawdown $d_t$ takes the value zero (since in that case $M_t=P_t$).

The role of $D$, 
\textbf{\emph{drawdown lookback}},
 is that it gives us flexibility not to be forced to look into the entire past when calculating drawdown, but instead simply look $D$ periods into the past. So drawdown at time $t$ is derived by comparing the portfolio value at time $t$ with the best (maximum)
  portfolio value over the preceding $D$ time periods. Using a time window for drawdown lookback has been seen previously in the literature, e.g.~in~\cite{zabarankin14}.

\subsection{Nonlinear formulation}
If we consider all possible time periods $t=1,\ldots, T$ then we would like drawdown to be small. There are a number of possibilities here, for example:
\begin{compactitem}
\item minimise average drawdown
\item minimise maximum drawdown 
\item minimise a weighted sum of maximum drawdown and average drawdown
\end{compactitem}

\noindent To proceed we shall first concentrate on minimising average drawdown. Let:
\begin{table}[H]
\begin{tabular}{p{1.3cm}p{13.8cm}}
$V_{it}$ & be the value (price) of asset $i$ at time $t$ \\
$A_{i}$  & be the  number of units of asset $i$ held in the current portfolio\\
$\delta_i$ & be the maximum proportion of the portfolio value at time $T$ that
 can be placed in asset $i$ \\
$C$ &  be the total value ($\geq 0$) of the current portfolio [$A_i$] at time $T$, $\sum_{i = 1}^N A_i V_{iT}$, plus 
cash change (either new cash to be invested or cash to be taken out) \\
$f_i^b$ & be the fractional cost of buying one unit of asset $i$ at time $T$, so that the cost incurred in buying one unit of
asset $i$ at time $T$ is  $f_i^b V_{iT}$ \\
$f_i^s$ & be the fractional cost of selling one unit of asset $i$ at time $T$, so that the cost incurred in selling one unit of
asset $i$ at time $T$ is  $f_i^s V_{iT}$ \\
$\gamma$ & be the limit ($0 \leq \gamma \leq 1$) on the proportion of $C$ that can be consumed by transaction cost \\
\end{tabular}
\end{table}
With respect to decision variables, let:
\begin{table}[H]
\begin{tabular}{p{1.3cm}p{13.8cm}}
$x_{i}$  & be the number of units ($\geq 0$, so we exclude shorting) of asset $i$ to be held in the portfolio \\
$G_i$ & be the transaction cost ($\geq 0$) associated with buying or selling asset $i$ at time $T$
\end{tabular}
\end{table}
Then our nonlinear formulation of the problem of minimising average drawdown is: 

\begin{equation}
\min \sum_{t = 1}^T d_t/T
\label{eq2}
\end{equation}

\noindent subject to equations~(\ref{eq1b}),(\ref{eq1}) and:

\begin{equation}
P_t = \sum_{i = 1}^N V_{it} x_i \;\;\; t = 1, \ldots, T
\label{eq4}
\end{equation}

\begin{equation} 
V_{iT}x_i
 \leq \delta_i P_T \;\;\; i=1,\ldots,N
\label{eq7}
\end{equation}

\begin{equation}
G_i \geq f_i^s (A_i - x_i) V_{iT} \,\,\,\, i = 1, \ldots, N
\label{jebeq4}
\end{equation}

\begin{equation}
G_i \geq f_i^b (x_i - A_i) V_{iT} \,\,\,\, i = 1, \ldots, N
\label{jebeq5}
\end{equation}

\begin{equation}
\sum_{i = 1}^N G_i \leq \gamma C
\label{jebeq6}
\end{equation}

\begin{equation}
P_T = C - \sum_{i = 1}^N G_i
\label{jebeq7}
\end{equation}

\begin{equation}
x_i, G_i \geq 0 \;\; i = 1, \ldots, N
\label{jebeq7a}
\end{equation}

\begin{equation}
d_t, P_t, M_t \geq 0 \;\;\; t=1,\ldots,T
\label{eq9}
\end{equation}

Equation~(\ref{eq2}) minimises average drawdown. 
Equation~(\ref{eq4}) defines the portfolio value variables ($P_t$). Clearly these variables can be eliminated by algebraic substitution, but we have left them in the formulation seen above for clarity of exposition.
 Equation~(\ref{eq7}) limits the proportion  of the portfolio value at time $T$  placed in any asset appropriately.
Equation (\ref{jebeq4}) defines the transaction cost associated with selling asset $i$, where we have sold the asset 
if the current holding $A_i$ is greater than the new holding $x_i$. 
Equation (\ref{jebeq5}) defines the transaction cost associated with buying asset $i$,
where we have bought the asset 
if the new holding $x_i$ is greater than the current holding $A_i$. 
Equation (\ref{jebeq6}) limits the total transaction cost.  
Equation (\ref{jebeq7}) is a balance constraint which ensures that the value of the portfolio after trading at time $T$ is equal to 
its value before trading (after accounting for the cash change, so $C$) minus the total transaction cost incurred.
Equations~(\ref{jebeq7a}),(\ref{eq9}) are the non-negativity constraints. 

The above formulation: optimise equation~(\ref{eq2}) subject to 
equations~(\ref{eq1b}),(\ref{eq1}),(\ref{eq4})-(\ref{eq9}), minimises average drawdown. Essentially in that formulation we are drawing on  historic asset price data $[V_{it}]$ over the period $t=1, \ldots, T$ to decide how we can best improve our existing portfolio of assets $[A_i]$, whilst taking account of any cash inflow/outflow, so as to minimise average drawdown over the period considered. 

In the computational results reported later we used asset sets $[i, i=1,\ldots,N]$ drawn from equity (stock) indices. Optionally, if so desired, we can include in the asset set a risk-free asset to represent investment in an interest-bearing ``cash'' account.

Note that there is one technical subtlety here - strictly the equations defining $G_i$ do not directly constrain $G_i$ to
be equal to the correct transaction cost (which is  $\mbox{max}[ f_i^s (A_i - x_i) V_{iT}, f_i^b (x_i - A_i) V_{iT}]$).
Rather $G_i$ is bounded below by the correct transaction cost (as the inequalities (equations~(\ref{jebeq4}),(\ref{jebeq5})) above indicate).
 We would expect that the numeric value for $G_i$ which we get from the optimisation to be equal to the  correct transaction cost found by direct calculation using 
the numeric $x_i$ value given by the optimisation, since otherwise we would have unallocated wealth (namely $G_i - \mbox{max}[ f_i^s (A_i - x_i) V_{iT}, f_i^b (x_i - A_i) V_{iT}]$) which is left untouched, so 
making no contribution to reducing portfolio drawdown. Computationally if  we wish to ensure that this situation never occurs then we simply amend the objective function (equation~(\ref{eq2})) to minimise $\Gamma \sum_{t = 1}^T d_t/T + \sum_{i = 1}^N G_i$ where $\Gamma$ is a large positive constant, thus ensuring that we minimise average drawdown as well the transaction cost variables $G_i$.

\subsection{Partial linearisation}

In our formulation  equations~(\ref{eq1b}) and~(\ref{eq1})  are nonlinear,  because of the presence of the maximisation term in equation~(\ref{eq1b}) and because of the division in equation~(\ref{eq1}). A \textbf{\emph{partial linearisation}} of this formulation can be made since the maximisation term, equation~(\ref{eq1b}), can be removed and replaced by the linear inequalities:
\begin{equation}
M_t \geq P_{\tau} 
\;\;\; 
\tau=t, t-1, \ldots, max[1,t-D] \;\;\;
t = 1, \ldots, T
\label{eq11}
\end{equation}

The formulation given above to minimise average drawdown then becomes: optimise equation~(\ref{eq2}) subject to 
equations~(\ref{eq1}),(\ref{eq4})-(\ref{eq11}).

To show that replacing nonlinear equation~(\ref{eq1b}) by linear equation~(\ref{eq11})  is valid is simple using a proof by contradiction. Suppose that when we optimise equation~(\ref{eq2}) subject to 
equations~(\ref{eq1}),(\ref{eq4})-(\ref{eq11}) we obtain, for some time period $t$, a value for $M_t$ than satisfies equation~(\ref{eq11}) but does not satisfy equation~(\ref{eq1b}). This can only occur if 
we have strict inequality, 
i.e.~$M_t > P_{\tau}  ~\tau=t, t-1, \ldots, max[1,t-D]$. 
Now equation~(\ref{eq1}) can be written as $d_t=100(1 - P_t /M_t)$
and we are minimising a function 
that involves $d_t$. So we would like $M_t$ as small as possible, 
consistent with the constraints upon it,
so as to make the negative contribution of the $P_t/M_t$ term in 
$d_t=100(1 - P_t /M_t)$
as great as possible.
The strict inequality for $M_t$ in relation to $P_{\tau}$ implies that we can reduce the numeric value for $M_t$, thereby reducing the value for $d_t$, and hence contradicting the assumption that we already had the optimal solution which involved optimising 
equation~(\ref{eq2}).  
Hence of the inequalities in 
equation~(\ref{eq11}) for time period $t$ one (or more) will be satisfied with equality at optimality, ensuring that
$M_t$ will indeed satisfy  equation~(\ref{eq1b}) and 
correspond to the appropriate maximum value over the  period $[ t, t-1, \ldots, max[1,t-D]]$. 

Note here that whether replacing nonlinear equation~(\ref{eq1b}) by linear equation~(\ref{eq11})  is valid  or not depends upon the particular optimisation problem involved. In other words equation~(\ref{eq11}) is not a general linearisation of equation~(\ref{eq1b}) that applies in all circumstances. For example if we were to maximise $\sum_{t = 1}^T d_t/T$ then this would be achieved by letting $M_t \rightarrow \infty$, which given the finite nature of the investment in each asset implied by equation~(\ref{jebeq7}) would mean that  whilst equation~(\ref{eq11}) was satisfied
equation~(\ref{eq1b}) would not be satisfied.

\subsection{Other objectives}
To minimise maximum drawdown we 
 introduce an artificial variable  $d^{max}$ ($\geq 0$) to represent maximum drawdown defined by:
\begin{equation}
d^{max} = max[~d_t ~|~ t = 1, \ldots, T]
\label{eqdmax}
\end{equation}

The formulation then is:
\begin{equation}
\min ~~~d^{max}
\label{eqobjmax}
\end{equation}

\noindent subject to equations~(\ref{eq1}),(\ref{eq4})-(\ref{eq11})  and:
\begin{equation}
d^{max} \geq d_t  \;\;\; t = 1, \ldots, T
\label{eqdmax1}
\end{equation}
\begin{equation}
d^{max} \geq 0 
\label{eq10}
\end{equation}

Equation~(\ref{eqdmax1}) is a linearisation of equation~(\ref{eqdmax}) using the same logic as used above in the linearisation of equation~(\ref{eq1b}).

 If $\lambda_1$ and $\lambda_2$ are the weights ($>0$) that we place on maximum drawdown and average drawdown respectively than to
minimise the weighted sum of maximum drawdown and average drawdown 
we:
\begin{equation}
\min ~~~\lambda_1 d^{max} + \lambda_2 \sum_{t = 1}^T d_t/T
\label{eqweight}
\end{equation}
subject to 
equations~(\ref{eq1}),(\ref{eq4})-(\ref{eq11}),(\ref{eqdmax1}),(\ref{eq10}).

In relation to a minor technical issue here  then, purely in the case when the objective is equation~(\ref{eqobjmax}), 
it is possible that 
when using equation~(\ref{eq11}) 
the numeric values assigned by the optimiser to the $M_t$ variables could be strictly greater than every element in the right-hand side of equation~(\ref{eq1b}).
However, since equation~(\ref{eq11}) is a $\geq$ constraint,  alternative optimal solutions exist where the $M_t$ values are artificially decreased to satisfy equation~(\ref{eq1b}).

Note here that we have  defined drawdown as the reduction in portfolio value  from the best portfolio value, expressed as a percentage of the best value ($M_t$) achieved, i.e.~$d_t = 100(M_t - P_t)/M_t$, equation~(\ref{eq1}). The approach given in this paper, including the partial linearisation given above, also applies (with only minor modifications) if (e.g.~as in~\cite{chekhlov05})  we define  drawdown
using $d_t = 100(M_t - P_t)/P_t$, i.e.~as the reduction in portfolio value  from the best portfolio value, expressed as a percentage of the portfolio value $P_t$.

\subsection{Computational considerations}

Readers familiar with the numeric solution of nonlinear programs will be aware that they are computationally much more challenging than solving linear programs. Computational benefit 
however
can sometimes
 be gained by imposing (non-trivial) bounds on decision variables. Limited computational experience indicated that for the solver 
(SCIP~\cite{scip}) we used this was indeed the case and so here we indicate the bounds we used. 

A valid upper bound for $x_i$ is  given by $\delta_iC/V_{iT}$, from equations~(\ref{eq7}),(\ref{jebeq7}). A valid upper bound for $G_i$ is $\gamma C$, from equation~(\ref{jebeq6}).
Valid lower and upper bounds for $P_t$ ($P_t^{min}$ and $P_t^{max}$ respectively) can be found by  optimising $P_t$ subject to 
equations~(\ref{eq4})-(\ref{eq9}). This is a simple linear program to solve, where we minimise $P_t$ to find the value for $P_t^{min}$, maximise  $P_t$ to find the value for $P_t^{max}$.

Valid lower and upper bounds for $M_t$ ($M_t^{min}$ and $M_t^{max}$ respectively) can be found by using $M_t^{min} = max[~P_{\tau}^{min} ~|~{\tau}=t, t-1, \ldots, max[1,t-D]]$ and $M_t^{max} = max[~P_{\tau}^{max} ~|~{\tau}=t, t-1, \ldots, max[1,t-D]]$. 
Valid lower and upper bounds for $d_t$ are given by 
$max[0, 100(1 - P_t^{max} /M_t^{min})]$ and 
$min[100, 100(1 - P_t^{min} /M_t^{max})]$.

The bounds that we used  therefore were:

\begin{equation}
x_i \leq \delta_iC/V_{iT}  \;\;\; i=1,\ldots,N
\label{eq15}
\end{equation}

\begin{equation}
G_i \leq \gamma C  \;\;\; i=1,\ldots,N
\label{eq15aa}
\end{equation}

\begin{equation}
P_t^{min}  \leq P_t \leq P_t^{max} \;\;\; t=1,\ldots,T
\label{eq15a}
\end{equation}

\begin{equation}
M_t^{min}  \leq M_t \leq M_t^{max} \;\;\; t=1,\ldots,T
\label{eq16}
\end{equation}

\begin{equation}
max[0, 100(1 - P_t^{max} /M_t^{min})]  
\leq d_t \leq min[100, 100(1 - P_t^{min} /M_t^{max})] \;\;\; t=1,\ldots,T
\label{eq18}
\end{equation}

Limited computational experience also indicated that for the nonlinear solver (SCIP) we used benefit could be gained if we replaced equality equation~(\ref{eq1}) by an inequality, as in equation~(\ref{eq1geq}) below:
\begin{equation}
d_t \geq 100(M_t - P_t) /M_t
\;\;\; t = 1, \ldots, T
\label{eq1geq}
\end{equation}
\noindent Replacing the equality definition of $d_t$ by this inequality definition can be shown to be valid using a very similar argument as was used above in terms of replacing  equality 
equation~(\ref{eq1b}) by inequality equation~(\ref{eq11}).

\subsection{Shorting}

In the formulation given above we have excluded shorting. The amendments necessary to include shorting are as outlined below.
For simplicity of exposition here we shall henceforth assume that all transaction costs are zero. Amending the formulation given previously above to incorporate the notational changes seen below relating to $x_i$ is relatively straightforward  (and is not given here for space reasons).  
Incorporating the transaction costs associated with shorting can be done in a  very similar manner as in the formulation above where we had transaction costs associated with buying/selling an asset (equations~(\ref{jebeq4})-(\ref{jebeq7a})).

Introduce $x_i^L, x_i^S$ as the number of units ($\geq 0$) of asset $i$ that we choose to hold in long/short positions respectively. Let $\delta_i^L$ and $\delta_i^S$ be the maximum proportions of the portfolio value at time $T$ that
 can be placed in asset $i$  in long/short positions respectively. Let $\Delta^L$ and $\Delta^S$ be the maximum overall proportions of the portfolio value at time $T$ that
 can be placed in long/short positions respectively. 

Then the decision variables $x_{i}$ are no longer non-negative, but are instead unrestricted in sign, and we have:

\begin{equation}
x_i = x_i^L - x_i^S \;\;\; i=1,\ldots,N
\label{s1}
\end{equation}

\begin{equation}
V_{iT}x_i^L \leq \delta_i^L P_T \;\;\; i=1,\ldots,N
\label{s2}
\end{equation}

\begin{equation}
V_{iT}x_i^S \leq \delta_i^S P_T \;\;\; i=1,\ldots,N
\label{s3}
\end{equation}

\begin{equation}
\sum_{i = 1}^N V_{iT} x_i^L \leq \Delta^L P_T
 \label{s4}
\end{equation}

\begin{equation}
\sum_{i = 1}^N V_{iT} x_i^S \leq \Delta^S P_T
 \label{s5}
\end{equation}

Equation~(\ref{s1}) relates the previously defined variable for the investment in asset $i$ to the new variables for investment in long/short positions.  
Equations~(\ref{s2}),(\ref{s3}) limit the proportion  of the portfolio value  placed in any asset at time $T$ appropriately, for both long and short positions.  Equations~(\ref{s4}),(\ref{s5}) limit the portfolio investment at time $T$  in long and short positions. 

In the presence of shorting then (although dependent on the values given to $\Delta^L$ and $\Delta^S$)
 it is possible that the in-sample portfolio value $P_t$ could potentially be negative in any time period $t$. 
Allowing the in-sample portfolio value to be in a loss position, even if by so doing we minimise some function of drawdown, would not in our view be desirable.  To avoid  this when shorting we simply retain equation~(\ref{eq9}) which ensures that $P_t$ is never negative. Hence, although we can short one or more assets, we never let the overall portfolio value go negative.


\section{Computational results}
\label{sec:results}

In this section we first discuss the data we used and the methodology adopted. We then present results for long-only drawdown portfolios, followed by result for portfolios with both long and short positions.

\subsection{Data and methodology}

We used three test instances drawn from major equity markets,  the EURO~STOXX~50, the FTSE~100 and the S\&P~500. For these markets we used daily price data over the period 2010-2016 (inclusive). The test instances we used have been manually curated to ensure that we know on any day the exact composition of the index. This means that when we come to rebalance the drawdown portfolio we \textbf{\emph{only}} consider for inclusion in the  portfolio assets that are in the index at the moment of rebalance. 
This means that our
results use no more information than was available
at the time, removing susceptibility to the influence
of survivor bias.

As noted in~\cite{meade11} most published analysis as to the size of transaction costs refers to US markets, and different estimates apply to equities with different capitalisations (companies of different sizes).
As our test instances involve a number of different equity markets (two of which are non-US) over a long time period then, 
given the difficulty in accurately estimating appropriate transaction costs, 
we assume in the results given below that transaction costs are zero.

The formulations discussed above that we investigated computationally were:
\begin{itemize}

\item formulation MINAVG: minimise average drawdown, 
optimise equation~(\ref{eq2}) subject to 
equations~(\ref{eq4})-(\ref{eq11}),(\ref{eq15})-(\ref{eq1geq})

\item formulation MINMAX: minimise maximum drawdown, 
optimise 
equation~(\ref{eqobjmax}) subject to 
equations~(\ref{eq4})-(\ref{eq11}),(\ref{eqdmax1}),(\ref{eq10}),(\ref{eq15})-(\ref{eq1geq})

\end{itemize}

We also examined these formulations, but with the addition of shorting (so these formulations, but modified as discussed above using equations~(\ref{s1})-(\ref{s5})). We denote the shorting formulations as MINAVG-S and MINMAX-S respectively.

Note here that although 
MINAVG, MINMAX, MINAVG-S and MINMAX-S
 are nonlinear programs the solver we used,
SCIP~\cite{scip}, is capable of solving them 
 to \textbf{\emph{proven global optimality}}~\cite{bussieck14, vigerske17, vigerske17a, vigerske18}. This is because SCIP restricts the type of nonlinear expression allowed. 
If, within the computational time limit allowed, SCIP cannot terminate with the proven optimal solution then it terminates with the best feasible solution found, but  as well provides a guaranteed percentage deviation from optimal for that solution.
For a technical explanation as to how SCIP can achieve proven global optimality see 
Vigerske and Gleixner~\cite{vigerske17a, vigerske18}.

We used an Intel Xeon @ 2.40GHz with 32GB of RAM and Linux as the operating system. The code was written in C++.

The methodology we adopted is successive periodic rebalancing over time. We start from
the beginning of our data set. We decide a portfolio using data taken from an in-sample period
corresponding to the first $T$ days. This portfolio is then held unchanged for a specified out-of-sample
period. We then rebalance (change) our portfolio, but now using the most recent $T$ days
 as in-sample data. The decided portfolio is then held unchanged for the specified out-of-sample period, and the process repeats until we have exhausted all of the data.

These results given below are for $T=30$ and $D=20$, so an in-sample time period of 30 (trading) days with a drawdown lookback of 20 (trading) days. We rebalanced the drawdown portfolio every 10 (trading) days, so an out-of-sample period of 10 days. The motivation for using low values for these parameters was that we were adopting a strategy of frequent rebalancing with a limited time horizon in terms of the immediate past. Such values seemed appropriate  for the context of our work (funds that seek to produce profit, or at least avoid loss, in the short-term; as discussed 
above). 

In terms of computation time we set a time limit for each rebalance of max(500, 7$N$) seconds. With some seven years of data the results shown below involve approximately 180 rebalances for each instance/case solved. We initialised the solution process using $C=1000$ and $A_i=0~i=1,\ldots,N$. At each rebalance we set $[A_i]$ equal to the portfolio $[x_i]$ decided at the last rebalance. This corresponds to a self-financing strategy, rebalancing with no cash inflow/outflow.

\subsection{Long-only drawdown portfolios}

For long-only  portfolios, so using formulations MINAVG and MINMAX, 
we examined two cases for the maximum proportion limit ($\delta_i$, equation~(\ref{eq7})), one where we limited the proportion to 0.1, so at most 10\% of portfolio value could be invested in any asset, and one where the proportion limit was 1, so potentially all of the investment could be in just a single asset. 

The results obtained are shown in Table~\ref{table2}.  In this table we (for space reasons)  also show results associated with shorting, and these will discussed later below.

To produce the first three in-sample columns in Table~\ref{table2} we first compute using the drawdown portfolio as decided by the optimiser, over the in-sample period associated with each rebalance: the average daily (logarithmic) return for that portfolio over the in-sample period; the maximum (\%) drawdown for that portfolio over the in-sample period; and the average (\%) drawdown for that portfolio over the in-sample period. Here drawdown is calculated as in 
equations~(\ref{eq1b}) and (\ref{eq1}), so involving drawdown lookback. The values shown in Table~\ref{table2} are then the averages for these three factors, as averaged over all rebalances. Note here that as we rebalance every 10 days the in-sample periods overlap.

The fourth in-sample column in Table~\ref{table2} gives the average computation time (per rebalance, in seconds) and the fifth and final in-sample column the percentage of optimal solutions found. As mentioned above although we are solving a nonlinear problem 
SCIP~\cite{scip}
 is able of solving problems such as MINAVG and MINMAX to proven global optimality and so this final in-sample column gives the percentage of rebalances for which the solution derived was proved to be optimal within the computational time limit imposed. 

The first three out-of-sample columns in Table~\ref{table2} have the same headings as the first three in-sample columns, but are computed using a \textbf{\emph{single time series of out-of-sample portfolio values}}. This single time series  of out-of-sample portfolio values is produced  by amalgamating together successive (non-overlapping) 10 day out-of-sample periods (one for each rebalance). For the instances shown in Table~\ref{table2} this single 
out-of-sample time series contained approximately 1800 daily   portfolio values.

The fourth out-of-sample column in Table~\ref{table2} shows the Sharpe ratio~\cite{sharpe66,sharpe75,sharpe94} annualised as in Pope and Yadav~\cite{pope94} using 252 trading days in a  year. Here, because we have used three different indices over a significant time period, we for ease of comparison use a risk-free rate of zero. 

The final out-of-sample column in Table~\ref{table2} shows the percentage of out-of-sample days in which the cumulative drawdown portfolio value  exceeded the index (when both were normalised to one at the start of out-of-sample period). With a normalisation to one at the start of out-of-sample period we are comparing cumulative return over time as achieved by the drawdown portfolio and the index. This statistic gives insight into the probability that the drawdown portfolio exceeds the index on any arbitrarily chosen day in the entire out-of-sample time period. 

For comparison purposes we also show in Table~\ref{table2} the values associated with the index, i.e.~the values for the factors tabulated that would be achieved were we to simply rebalance to the index portfolio every 10 days.

For the EURO~STOXX~50 we can see from Table~\ref{table2} that in-sample all four drawdown portfolios have superior return performance, as well as superior maximum and average drawdown performance when compared to the index itself. It should be emphasised here that although we restrict the set of assets that can be in the drawdown portfolio to the assets that are in the index (for consistency of comparison) the MINAVG and MINMAX approaches that we have formulated above make no use of index values nor of any knowledge as to how the index value is computed (in terms of a weighted sum of index asset prices).

Out-of-sample 
all four drawdown 
 portfolios effectively dominate the index in terms of (cumulative) return achieved. Numerically for over 99\% of the out-of-sample days the drawdown portfolios have a cumulative return that exceeds that of the index. 
With respect to drawdown out-of-sample three of the four drawdown portfolios dominate the index with respect to both maximum and average drawdown. Sharpe ratios for all four of the drawdown portfolios are better than the Sharpe ratio for the index.

\begin{figure}[!htb]
\begin{center}
\caption{Out-of-sample performance: EURO~STOXX~50}\label{fig2}
\includegraphics[width=0.9\textwidth]{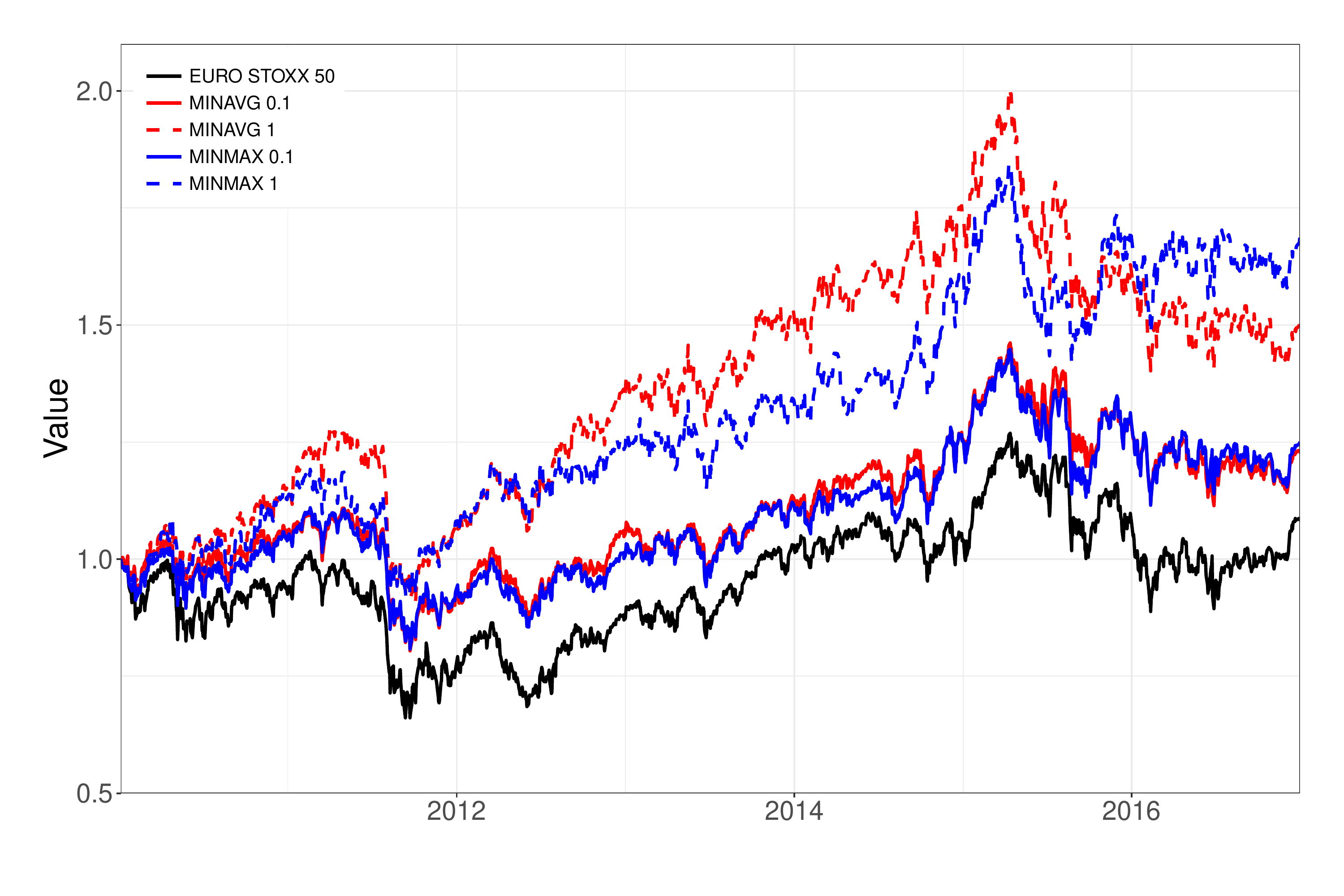}
\end{center}

\end{figure}

Figure~\ref{fig2} shows graphically, for the
EURO~STOXX~50,  
 the out-of-sample performance for each of the MINAVG and MINMAX portfolios summarised in Table~\ref{table2}, as well as the performance of the index (all values normalised to one at the start of the out-of-sample period).

One item of note for the EURO~STOXX~50 is the significant difference between the maximum drawdown values in-sample and out-of-sample 
(e.g.~3.41\% and 19.51\% respectively for MINAVG~0.1). This arises because the in-sample maximum drawdown is an average over approximately 180 rebalances of 180 maximum drawdowns, each computed from a time series containing 30 values; whereas the maximum drawdown out-of-sample is a single value computed for a time series containing approximately 1800 values. As such these maximum drawdown values are not directly comparable.

For the FTSE~100 we can see from Table~\ref{table2} that in-sample all four drawdown portfolios have superior return performance, as well as superior maximum and average drawdown performance when compared to the index itself. 
Out-of-sample Table~\ref{table2} 
shows that three of the four drawdown portfolios effectively dominate the index in terms of (cumulative) return achieved. 
With respect to drawdown out-of-sample two of the four drawdown portfolios (MINAVG~0.1 and MINMAX~0.1) dominate the index with respect to both drawdown measures. Sharpe ratios for all four of the drawdown portfolios are better than the Sharpe ratio for the index.

For the S\&P~500 we can see from Table~\ref{table2} that in-sample all four drawdown portfolios have superior return performance, as well as superior maximum and average drawdown performance when compared to the index itself. 

Of particular note for the S\&P~500 is the high percentage of optimal solutions obtained (higher than the corresponding cases for either the EURO~STOXX~50 or the FTSE~100).
For the S\&P~500 we have the number of assets $N$ equal to 500, much larger that the values of $N$ for the EURO~STOXX~50 ($N=50$) or FTSE~100 ($N=100$). Although a larger value for $N$ increases the number of decision variables and linear constraints the number of nonlinear constraints depends only upon $T$ 
(see equation~(\ref{eq1})), and all the cases examined in Table~\ref{table2} have $T=30$. The reason why we have a much higher percentage of optimal solutions found for the S\&P~500 is, we believe, due to the fact that for this index the growth over the period is much larger than for either the EURO~STOXX~50 or the FTSE~100 (compare the average return  on the index given in 
Table~\ref{table2} for the S\&P~500 with that for the  
EURO~STOXX~50 or the FTSE~100).
If an index is growing then one might reasonably suppose that it is easier to find a portfolio with drawdown zero, or close to zero 
(i.e.~a portfolio that effectively grows for all/most of the in-sample period). The very low average in-sample drawdown values seen in Table~\ref{table2} for the S\&P~500 support this argument that because of the growth in the index optimal drawdown portfolios will be more easily found.

For the S\&P~500 we  have from Table~\ref{table2} that out-of-sample all four drawdown portfolios dominate  the index in terms of maximum and average drawdown, as well as in terms of the Sharpe ratio. 
Over the (approximately 1800 day) 
out-of-sample period all of the drawdown portfolios 
exceed the index in terms of cumulative return for over 89\% of  days. 

The average values in Table~\ref{table2} show that all of the drawdown portfolio cases examined dominate the index in terms of return,  maximum drawdown and average drawdown both in-sample and out-of-sample.
 Sharpe ratios for all four of the drawdown portfolios are better than the corresponding Sharpe ratios for the indices.
 Of particular note here is the performance of the MINMAX~0.1 portfolio which out-of-sample has the highest  return and  the lowest maximum drawdown with the associated drawdown portfolios exceeding the index for 99.6\% of days in the out-of-sample period.

\subsection{Long and short  drawdown portfolios}

For drawdown portfolios involving shorting, so using formulations MINAVG-S and MINMAX-S, 
we set the maximum proportion limits ($\delta_i^L$ and  $\delta_i^S$ equations~(\ref{s2}),(\ref{s3})) to 0.1, so at most 10\% of portfolio value could be invested (long or short) in any asset. We used $\Delta^S =0.1$ (see equation~(\ref{s5}))  so the
maximum overall proportion of the portfolio value at time $T$ that
 could be placed in short positions (across all assets) was at most 10\%.
Due to the nature of shorting (where short positions generate funds to purchase assets) we typically have  $\Delta^L = 1 + \Delta^S$, so here we used  
$\Delta^L=1.1$ (see equation~(\ref{s4})).

Comparing the MINAVG-S~0.1 and MINMAX-S~0.1 results with shorting to the results for MINAVG~0.1 and MINMAX~0.1 without shorting we have that in-sample for all three instances, the average daily return, maximum drawdown and average drawdown are all superior when shorting is allowed as compared with the results without shorting. Allowing shorting provides additional flexibility in choosing the drawdown portfolio and this clearly results for our instances in better in-sample performance.

Out-of-sample the picture is more mixed. 
For the EURO~STOXX~50 the maximum and average drawdown results are better out-of-sample when shorting is allowed than the equivalent cases without shorting. However the return achieved is much less.

\begin{figure}[!htb]
\begin{center}
\caption{Out-of-sample performance with shorting: EURO~STOXX~50}\label{fig5}
\includegraphics[width=0.9\textwidth]{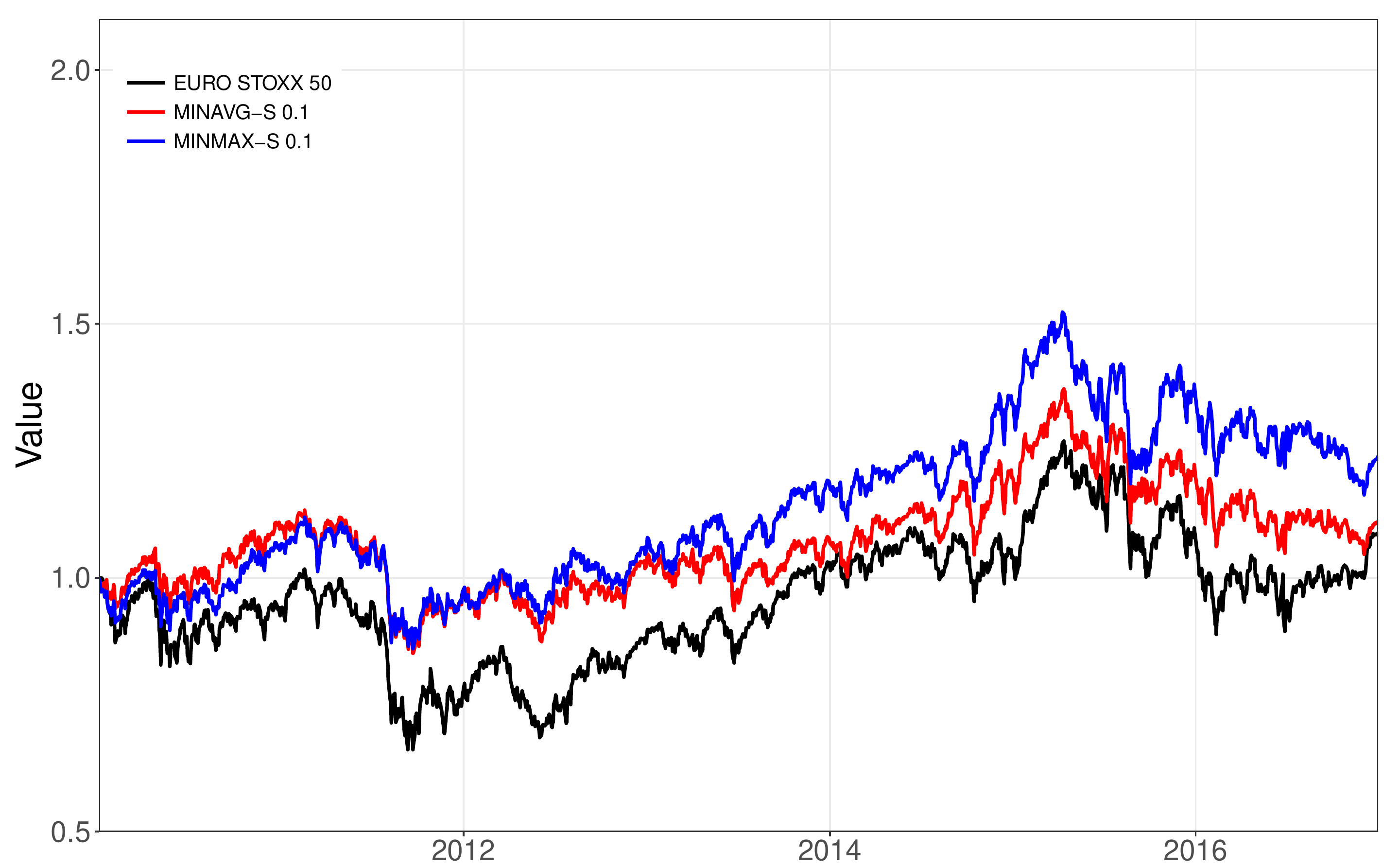}
\end{center}

\end{figure}

Figure~\ref{fig5} shows graphically, for the
EURO~STOXX~50, 
 the out-of-sample performance for each of the MINAVG-S and MINMAX-S portfolios summarised in Table~\ref{table2}, as well as the performance of the index (all values normalised to one at the start of the out-of-sample period).

For the FTSE~100 only maximum drawdown is better out-of-sample than the equivalent cases without shorting. For the S\&P~500 it also seems clear that (on balance) allowing shorting does not pay off in terms of improved out-of-sample performance.

Comparing the results for MINAVG-S~0.1 and MINMAX-S~0.1 with shorting to the associated index values we have that for all three instances the out-of-sample shorting results correspond to portfolios that are superior to the index in terms of maximum drawdown, average drawdown and Sharpe ratio. For the EURO~STOXX~50 and the FTSE~100 the drawdown portfolios provide an average daily return that exceeds that of the index. For the S\&P~500 the returns provided fall just below that of the index, although note that, as occurred with the non-shorting portfolios, we have that over the (approximately 1800 day) out-of-sample period the drawdown portfolios exceed the index in terms of cumulative return for over 89\% of  days.

Although obviously dependent on the instances examined, as well as the various parameter values adopted (such as for $T$, $D$, 
$\delta_i$, $\delta_i^L$, $\delta_i^S$,  $\Delta^L$ and  $\Delta^S$), these results would indicate that long-only drawdown portfolios would be preferable to drawdown portfolios that allow shorting. However we would note that the drawdown portfolios with shorting that we examined still effectively dominate the index.

\section{Conclusions}
\label{sec:con}
In this paper we have considered the problem of minimising drawdown in a portfolio of financial assets. 
We discussed how the work presented in this paper follows the general pattern of dealing with a financial portfolio optimisation problem that first appeared in the financial literature  now being considered from an Operational Research perspective.

We formulated the problem (minimising either average drawdown, maximum drawdown, or a weighted combination of the two) as a nonlinear program and showed how it can be partially linearised  by replacing one of the nonlinear constraints by equivalent linear constraints. 
Our formulation incorporated cash inflow/outflow, and can be used either to create an initial portfolio from cash or to rebalance an existing portfolio. Transaction costs associated with buying or selling an asset were included.

Computational results were presented (generated using the nonlinear solver SCIP) for three test instances drawn from the EURO~STOXX~50, the FTSE~100 and the S\&P~500 with daily price data over the period 2010-2016. 
We considered long-only drawdown portfolios as well as portfolios with both long and short positions.

Our results showed that (within the computational time limits imposed) we were able to find globally optimal minimal drawdown portfolios for a significant percentage of the cases examined. Our results indicated that (on average) our minimal drawdown portfolios dominated the market indices in terms of return, Sharpe ratio,     maximum drawdown and average drawdown
 over the (approximately 1800 trading day) out-of-sample period. 

Finally we would note here that we believe that the work presented in this paper indicates the value of adopting an Operational Research perspective on a financial portfolio optimisation problem.

\begin{landscape}


\begin{table}[!htb]
\centering
 {\scriptsize
\caption{Computational results}
\label{table2}
\begin{tabular}{lcc|ccccc|ccccc}\cline{1-13} 
Instance & Case & Proportion & \multicolumn{5}{c|}{In-sample} &
\multicolumn{5}{c}{Out-of-sample}  \\
 & & limit & Average & Maximum & Average  
& Average & Optimal
& Average & Maximum & Average & Annualised & Percentage of days \\
& &  & daily & drawdown & drawdown 
& computation & solutions &
daily & drawdown & drawdown & Sharpe & cumulative portfolio \\
& & & return & (\%) & (\%) 
& time (secs) & found (\%) &
return & (\%) & (\%) & ratio & value exceeds index  \\
\cline{1-13} 
EURO~STOXX~50 & Index & &
 
0.000074	&	6.60	&	2.38	 
&
 & &
0.000048	&	22.32	&	3.19	 
& 0.055
\\
& MINAVG & 0.1 &
 
0.001652	&	3.41	&	0.84	 
&

279.4	&	52.8	&	 

0.000116	&	19.51	&	2.55	 
& 0.162
 & 99.6
\\
& MINAVG & 1 &
 0.002164	&	2.62	&	0.55	
&
 
218.6	&	67.2	&	 

0.000229	&	24.11	&	2.60	 

& 0.314
& 99.8
\\
& MINMAX & 0.1 &
 
0.001288	&	2.93	&	0.98	 
&

354.8	&	31.1	&

0.000125	&	19.52	&	2.59	
& 0.177
& 99.3
\\
& MINMAX & 1 &
 
0.001825	&	2.04	&	0.72	 
&

340.5	&	32.8	&	 

0.000292	&	18.39	&	2.60	
& 0.405
& 99.4
\\
 &  &  &  &  &  &  &  & \\
& MINAVG-S & 0.1 &
 
0.002176	&	2.69	&	0.57	&	
 
273.8	& 	62.8	 
	 
&
 
0.000058	&	17.34	&	2.50	&	0.084	&	99.6	 
\\

& MINMAX-S & 0.1 &
 
0.001717	&	2.17	&	0.72	&	
 
292.9	& 	51.1	 
	 
&
 
0.000120	&	17.95	&	2.49	&	0.179	&	99.2	 
\\
 &  &  &  &  &  &  &  & \\

FTSE~100 & Index & &
 
0.000170	&	4.95	&	1.68	 
&
  & &
0.000148	&	15.63	&	2.22	 
& 0.232

\\
& MINAVG & 0.1 &
 
0.002379	&	1.38	&	0.21	 
&

214.5	&	75.7	&	 

0.000296	&	11.99	&	1.79	 
& 0.561
& 95.2
 \\
& MINAVG & 1 &
 
0.002602	&	1.01	&	0.14	 

&

87.50	&	93.2	&	 

0.000143	&	18.67	&	2.18	 
& 0.245
& 67.1
\\
& MINMAX & 0.1 &

0.002092	&	0.85	&	0.27	 

&
 
497.2	&	32.2	&

0.000337	&	11.07	&	1.80	 
& 0.644
& 99.4
 \\
& MINMAX & 1 &
 
0.002266	&	0.63	&	0.22	 
&

437.2	&	41.8	&	 

0.000208	&	18.62	&	2.04	
& 0.358 
& 92.0
 \\
 &  &  &  &  &  &  &  & \\

& MINAVG-S & 0.1 &
 
0.002966	&	0.79	&	0.09	&	
 
135.2	& 	81.9	 
	 
&
 
0.000295	&	11.53	&	1.83	&	0.553	&	98.7	 
\\

& MINMAX-S & 0.1 &
 
0.002605	&	0.65	&	0.26	&	
 
396.8	& 	47.5	 
	 
&
 
0.000269	&	10.13	&	1.88	&	0.517	&	98.3	 
\\
 &  &  &  &  &  &  &  & \\
S\&P~500 & Index & &
 
0.000435	&	4.18	&	1.36	 
&
  & &
0.000387	&	16.77	&	1.79	 
& 0.626
\\
& MINAVG & 0.1 &
 
0.003178	&	0.43	&	0.05	 
&

292.3	&	93.8	&	 

0.000357	&	10.29	&	1.63	 
& 0.712
& 93.3
\\
& MINAVG & 1 &
 
0.003442	&	0.19	&	0.02	 

&

170.7	&	97.2	&	 

0.000334	&	11.57	&	1.68	 
& 0.655
& 90.6
 \\
& MINMAX & 0.1 &
 
0.003006	&	0.26	&	0.09	 

&

856.1	&	78.5	&	 

0.000421	&	10.06	&	1.65	 
& 0.828
& 100
\\
& MINMAX & 1 &
 
0.003283	&	0.09	&	0.03	 
&

527.2	&	87.6	&	 

0.000347	&	9.50	&	1.59	 
& 0.705
& 89.7
\\
 &  &  &  &  & &  &  & \\

& MINAVG-S & 0.1 &
 
0.003684	&	0.08	&	0.01	&	
 
137.2	& 	100	 
	 
&
 
0.000360	&	10.92	&	1.74	&	0.694	&	93.5	 
\\

& MINMAX-S & 0.1 &
 
0.003270	&	0.04	&	0.01	&	
 
375.1	& 	93.2	 
	 
&
 
0.000355	&	8.71	&	1.52	&	0.738	&	89.7	 

\\
&  &  &  &  & &  &  & \\
Average & 	Index 	&	& 0.000226 &	5.24 &	1.81	& & 	&	0.000194	& 18.24 & 	2.40	& 0.304	   \\
&	MINAVG & 0.1	&	0.002403	& 1.74 &	0.37 &	262.1 &	74.1	 & 0.000256	& 13.93	& 1.99	& 0.478 & 96.0	   \\
& 	MINAVG & 1	&	0.002736 &	1.27 &	0.24	 & 158.9 &	85.9 &	0.000235	 & 18.12 &	2.15 & 0.405  &	85.8	   \\
&	MINMAX  & 0.1	&	0.002129 &	1.35 &	0.45 &	569.4	 & 47.3 &	0.000294 &	13.55 &	2.01 &	0.550 & 99.6	   \\
&	MINMAX & 1		& 0.002458 &	0.92	& 0.32 &	435.0 &	54.1	& 0.000282	& 15.50 &	2.08 &	0.489 & 93.7	\\
  &  &  &  & &  &  & & \\
 & MINAVG-S & 0.1 &

0.002942 &	1.19	& 0.22	& 136.2 & 91.0 &	0.000238 & 	13.26 	& 2.02 &	0.444 &	97.3	 \\  
	
& MINMAX-S & 0.1 &
0.002531	& 0.95 &	0.33 & 386.0 & 70.4 &	0.000248	& 12.26	& 1.96 &	0.478 &	95.7	  \\
 &  &  &  &  &  &  & & \\
\cline{1-13} \\
\end{tabular}
}
\end{table}


\end{landscape}

 \clearpage
\newpage
 \pagestyle{empty}
\linespread{1}

\end{document}